\documentclass[letterpaper, 10 pt, conference]{ieeeconf}  

\IEEEoverridecommandlockouts                              

\usepackage{subcaption}
\usepackage{graphicx}
\usepackage{hyperref}
\usepackage{tabularx}
\usepackage{multirow}
\let\labelindent\relax
\usepackage{enumitem}
\usepackage{balance}
\usepackage{booktabs}
\usepackage{cuted}
\usepackage{enumitem}
\usepackage{tikz}
\usetikzlibrary{positioning,arrows.meta,shapes,fit,calc}
\usepackage{subcaption}

\usepackage{xspace}
\newcommand{\dataset}{CCO\xspace}

\title{\LARGE \bf 
How to start a conversation proactively?
}

\author{Mohit Nathani, Aaditya Gupta, Eshwar Dhande, and Chayan Sarkar\\
TCS Research, India\\
}

\begin{document}

\maketitle
\thispagestyle{empty}
\pagestyle{empty}


\begin{abstract}
Proactively initiating a conversation is a fundamental social competence and a critical capability for social robots. However, determining how and when to start a conversation in a given situation remains an open challenge, primarily due to the lack of a principled definition of conversational context. While recent large language models (LLMs) demonstrate impressive conversational fluency, they are predominantly turn-reactive and often produce generic or monotonous openers when tasked with initiating interaction, even when contextual information is available. Such limitations are particularly problematic for embodied social robots, where inappropriate or poorly timed initiations can negatively affect user engagement and rapport.
In this work, we propose a schema of contextual variables that jointly characterize situations relevant for proactive conversation initiation. The schema is designed to structure perceptual, situational, and interactional cues such that they can be effectively exploited by language models. Furthermore, motivated by the absence of suitable training data, we introduce a curated dataset of contextually-aligned conversation openers (CCO)~\footnote{https://github.com/mohit-nathani1/conversation-opener-dataset} covering diverse everyday scenarios. Using this dataset, we fine-tune small language models (SLMs) to generate proactive conversation initiations.
Through a comparative evaluation involving zero-shot LLMs, off-the-shelf SLMs, and our fine-tuned models, we demonstrate that explicit contextual structuring and targeted data curation substantially improve the quality and contextual appropriateness of conversation openers. Our results highlight the importance of context modeling for proactive social interaction and provide actionable building blocks for conversational robots.

\end{abstract}


\section{Introduction}

Conversational artificial intelligence has advanced rapidly with the emergence of large language models (LLMs), which exhibit strong fluency and contextual coherence across domains. These systems excel at sustaining dialogue in response to user input. However, most contemporary conversational architectures generate utterances only after being addressed (reactive)~\cite{deng2025proactive, pramanick2022conversational}. The problem of autonomous conversation initiation, deciding when and how a system should begin interaction in socially complex settings, remains comparatively underexplored~\cite{liu2025proactiveconversationalagentsinner}.

Initiating a conversation is not merely producing the first utterance; it involves selecting what to say and how to say it in a manner appropriate to situational, relational, and temporal context. An utterance may be grammatically correct yet socially inappropriate if it fails to account for interactional sensitivity, tone, or timing. Thus, generating conversational openers is not only a language generation problem, but a form of context-dependent social decision-making.

\begin{figure}
    \centering
    \includegraphics[width=0.99\linewidth]{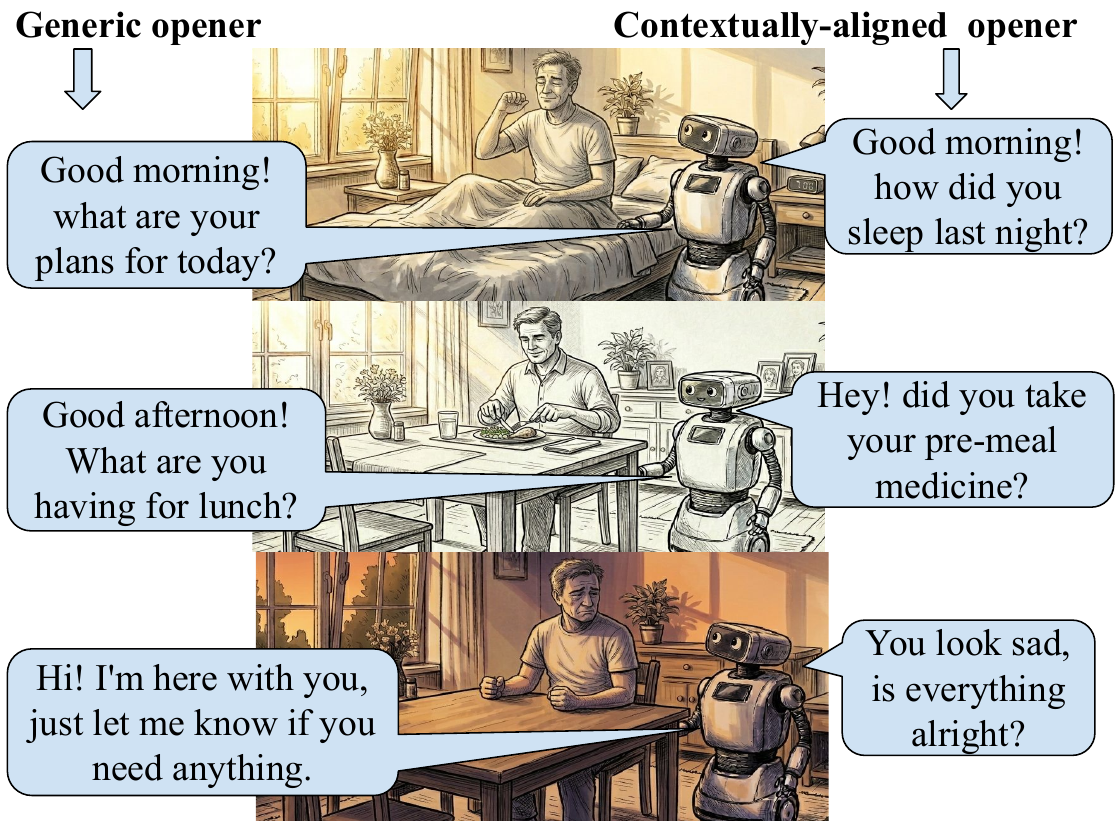}
    \caption{Illustration of conversational openers at different times of day, contrasting generic with contextually-aligned that incorporate user state, routine, and situational context.}
    \label{fig:illustration}
    \vspace{-2em}
\end{figure}

In human interaction, conversation initiation relies on implicit reasoning over contextual cues, communicative intent, and social norms. Speakers balance informational goals with potential social cost, adapting phrasing to the perceived situation and relationship. Modeling this balance computationally remains a central challenge for socially aware conversational agents (Fig.~\ref{fig:illustration}).

Early approaches relied on rule-based conversational systems, where predefined mappings associate interaction states with specific utterances or strategies~\cite{SINGH20228852}. While such systems offer interpretability and control, they scale poorly as interactional complexity grows. The combinatorial growth of contextual conditions leads to brittle designs, and rigid rules fail to capture nuanced variation in phrasing and tone. LLMs provide an alternative through conditional generation over structured inputs~\cite{sarkar2026tool}. However, whether they can reliably prioritise contextual constraints and initiate conversations autonomously remains an open question.

Despite their flexibility, LLMs exhibit limitations in proactive interaction. They often produce repetitive or generic openers under loosely constrained or recurring contexts~\cite{wu2026rise}, reducing naturalness and causing user fatigue in long-term interactions. Without explicit mechanisms for diversity or memory-aware adaptation, LLM-generated proactive utterances may appear formulaic or intrusive, undermining socially appropriate engagement.

Moreover, deploying LLMs for continuous proactive reasoning raises practical and ethical concerns. High computational cost, inference latency, and cloud dependence limit their suitability for embodied agents in privacy-sensitive contexts. In contrast, small language models (SLMs) can be deployed locally on robots or edge devices, enabling low-latency, privacy-preserving interaction while processing contextual signals on-device~\cite{nathani2024knowledge}.

In this work, we investigate language models as structured social decision-makers for proactive conversation initiation. Our contributions are threefold:
\begin{itemize}
    \item We formalize a schema of interaction variables for socially appropriate conversation initiation.
    \item We introduce a human-verified dataset mapping structured social contexts to calibrated conversation openers.
    \item We fine-tune SLMs and evaluate them against LLMs and their pre-trained counterparts, showing that locally deployable models achieve improved performance in socially grounded initiation.
\end{itemize}

\section{Related Work}

\subsection{Conversational Dialogue Systems}
Research on conversational systems has evolved from rule-based pipelines (NLU, dialogue state tracking, policy learning, NLG) to large-scale neural models capable of open-domain interaction \cite{yi2025survey}. Pre-trained language models such as DialoGPT \cite{zhang-etal-2020-dialogpt} demonstrate strong performance in multi-turn dialogue, while systems like BlenderBot~3 incorporate long-term memory, retrieval, and safety mechanisms \cite{unknown}. 

In parallel, task-oriented dialogue research has explored end-to-end generative formulations~\cite{pramanick2019enabling,pramanick2022robotic}. SimpleTOD~\cite{10.5555/3495724.3497418} and SOLOIST~\cite{peng-etal-2021-soloist} unify belief tracking and generation within autoregressive models, reducing reliance on explicit policies. More recent work shows that structured behaviour can emerge without explicit state annotations \cite{zhao2022unids}. Despite these advances, most systems remain primarily designed for reactive response generation rather than autonomous conversation initiation.




\subsection{Proactive and Mixed-Initiative Interaction}
Beyond purely reactive dialogue, mixed-initiative systems investigate how agents can dynamically contribute to conversations by initiating turns, asking questions, or offering suggestions. Deng et al.~\cite{deng2023knowledgeenhancedmixedinitiativedialogueemotional} introduce a knowledge-enhanced mixed-initiative framework for emotional support dialogues, demonstrating that controlled initiative improves conversational depth and user outcomes.

In human–robot interaction, FlowAct~\cite{dhaussy2024flowact} monitors multimodal contextual cues to trigger proactive robot behaviors, emphasizing the importance of environmental sensing for initiative timing. Qiu et al.~\cite{qiu-etal-2020-structured} explore latent conversational structure using unsupervised models, revealing underlying turn-taking and initiative flows. While these approaches model initiative dynamics, they do not explicitly address how to linguistically formulate socially appropriate conversation openers across diverse contexts.

\subsection{Social and Context-Aware Dialogue Systems}
Socially aware dialogue systems incorporate user traits, emotional state, and situational context to produce interactionally appropriate responses. Prior work has shown that socially expressive responses increase user satisfaction and trust compared to purely task-oriented communication~\cite{wang2025task}. Persona-based models~\cite{liu2020impressmedialoguegeneration} and emotion-conditioned generators~\cite{DBLP:journals/corr/abs-1811-10990} demonstrate the importance of modeling stable user attributes and affective cues in conversational behavior.

Recent work has also explored grounding conversational behavior in visual context. For instance, \cite{janssens2024integrating} investigate generating socially appropriate conversation-starting questions from images, comparing pipeline approaches with end-to-end vision-language models and large language models. Reinforcement learning approaches explicitly model trade-offs between agent autonomy and social acceptability~\cite{Kraus_2023}, highlighting that effective proactive dialogue requires balancing initiative with relational sensitivity. However, most social-context-aware systems focus on response adaptation rather than autonomous conversation initiation.

\begin{figure*}
    \centering
    \includegraphics[width=0.95\linewidth]{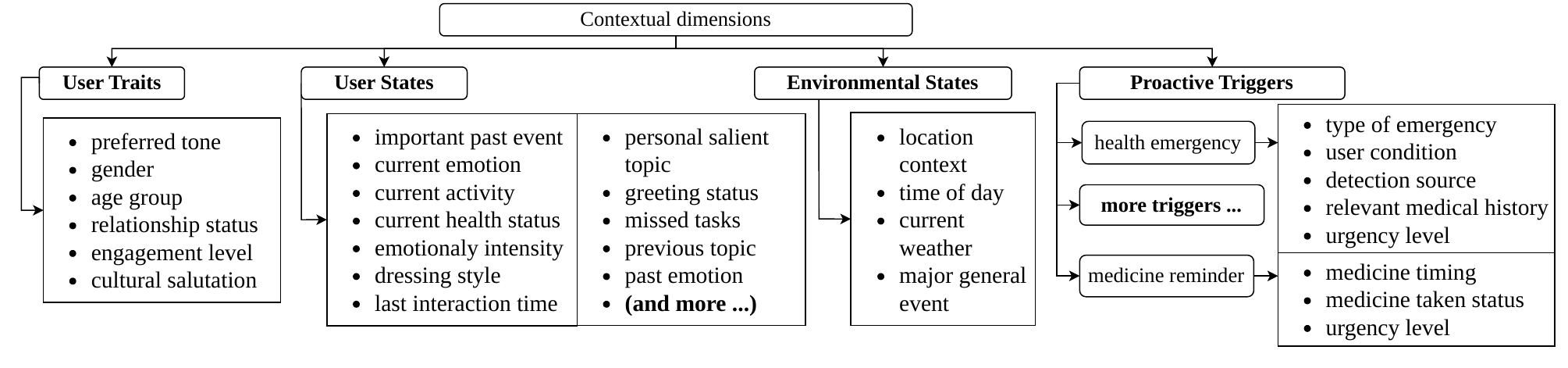}
    \caption{Dimensions of contextual schema for generating aligned conversational openers.}
    \label{fig:schema}
    \vspace{-1em}
\end{figure*}

\subsection{Theoretical Foundations of Social Interaction}
Foundational theories from social psychology and sociology provide important grounding for proactive conversational behavior. Empirical studies show that individual traits, such as extraversion and openness, systematically influence conversational dynamics including topic diversity and linguistic alignment~\cite{Fisher_2024}. Sociolinguistic research further demonstrates that conversational entry points vary with age, power relations, familiarity, and cultural norms~\cite{yuwono-santoso-2024-address}.

Complementing these perspectives, the Persuasive Systems Design framework~\cite{article} emphasizes that interactive systems must adapt strategies according to user characteristics and situational context, reinforcing the need for structured representations of social variables. Together, these theories motivate the explicit modeling of contextual and relational factors for computational conversation initiation.

\subsection{Positioning of This Work}
In contrast to prior work on reactive dialogue and limited initiative-taking, we focus explicitly on proactive conversation initiation. By formalising interaction context through a structured schema and training models to generate socially calibrated openers, our approach bridges dialogue systems and computational models of social decision-making.

\section{Conversation Context Schema}

To support proactive and socially appropriate conversation initiation, we represent interaction context through a structured schema of variables. The framework decomposes interaction into four components --- (i) user traits, (ii) user states, (iii) environmental states, and (iv) proactive triggers---enabling explicit reasoning about what to say and when and why to initiate interaction (Fig.~\ref{fig:schema}).

\textbf{User traits} capture relatively stable characteristics that influence conversational style, such as preferred tone, greeting form, verbosity, and initiative level. These variables guide stylistic realization rather than initiation decisions, enabling consistent personalization across sessions. Identity-linked attributes (e.g., age group) are used conservatively to ensure appropriate language complexity and respectful addressing~\cite{yuwono-santoso-2024-address, zhang2025personalizing}.

\textbf{User states} represent the user’s immediate condition, including activity, emotional state, and health status. These variables regulate interruption sensitivity, empathy calibration, and conversational readiness, allowing the system to avoid poorly timed initiation and generate contextually appropriate openers.

\textbf{Environmental states} capture external situational context such as time of day, location, weather, and shared events. These signals ground conversation initiation in the surrounding environment, supporting appropriate timing and natural small talk aligned with the user’s context.

\textbf{Proactive triggers} enable event-driven conversation initiation for recurring assistance scenarios, including medication adherence, health checkups, exercise, hydration, and safety-critical events. Each trigger follows a shared interaction logic comprising (i) prerequisite checks, (ii) contextual constraints, and (iii) goal-state verification, ensuring that initiation is timely, non-redundant, and contextually valid. For example, medication reminders are generated only when temporal constraints are satisfied and the dose has not already been taken. Small talk functions as a relational trigger, leveraging contextual signals to support engagement when no urgent task is active.

Across all components, variables are categorized into five functional roles: modifiers, context, constraints, prerequisites, and goal states. Modifiers influence stylistic realization, context variables provide situational grounding, constraints regulate when initiation is appropriate, prerequisites capture necessary preconditions, and goal states represent task completion or progress signals.

\begin{figure*}
    \centering
    \includegraphics[width=0.95\linewidth]{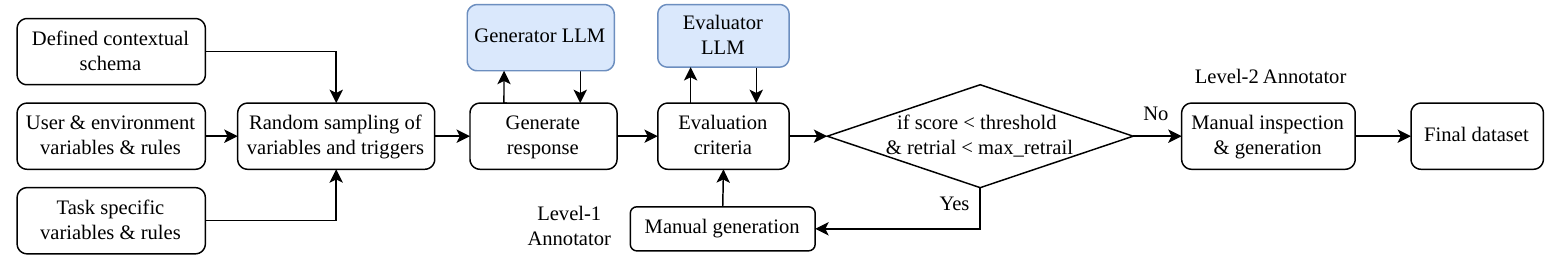}
    \caption{Our data generation pipeline composed of a generator LLM, evaluator LLM, and manual intervention.}
    \label{fig:data_gen_pipeline}
\end{figure*}

\section{Dataset}

We introduce the \textbf{Contextually-aligned Conversation Opener (\dataset)} dataset. The task is formulated as conditional generation: given a structured representation of a social scenario, a model must produce a single-sentence conversational opener that is socially appropriate and contextually grounded. Each \dataset instance consists of (i) a high-level \textit{trigger} specifying conversational intent and (ii) a structured context comprising user state, user traits, environmental state, memory state, and trigger-specific variables. The target output is a short, natural language opener. \dataset is designed to support training and evaluation of small language models (SLMs) for proactive, privacy-preserving conversational agents.

\begin{table*}[t]
\centering
\small
\begin{tabular}{lcccccccc}
\toprule
\textbf{Split} & \textbf{Small Talk} & \textbf{Medicine} & \textbf{Hydration} & \textbf{Exercise} & \textbf{Checkup} & \textbf{Env. Emerg.} & \textbf{Health Emerg.} & \textbf{Total} \\
\midrule
Train & 281 & 285 & 302 & 253 & 286 & 177 & 176 & 1760 \\
Validation & 42 & 35 & 48 & 40 & 32 & 24 & 24 & 245 \\
Test & 83 & 84 & 56 & 95 & 84 & 40 & 40 & 482 \\
\midrule
\textbf{Total} & 406 & 404 & 406 & 388 & 402 & 241 & 240 & 
\textbf{2487} \\
\bottomrule
\end{tabular}
\caption{Distribution of samples across dataset splits and trigger categories in \dataset.}
\label{tab:dataset_distribution}
\vspace{-2em}
\end{table*}

\subsection{Data Construction and Quality Control}

\dataset is constructed using a hybrid pipeline combining structured scenario sampling, LLM-based generation, automated evaluation, and multi-stage human refinement (Fig.~\ref{fig:data_gen_pipeline}). We first define a comprehensive contextual schema spanning state variables and trigger-specific attributes. Context instances are generated by sampling variable combinations, including both typical and edge-case configurations to ensure diversity and robustness.

Each sampled context is provided to a generator LLM (GPT-5-mini), which produces a candidate opener conditioned on the full structured schema. Responses are evaluated using an LLM-as-a-Judge framework~\cite{lin2023llm} across six qualitative dimensions on a 0–5 scale. Based on the average score, samples are routed through a two-stage annotation pipeline.

Samples with an average score $\geq$ 4.0 are directly sent for Level-2 validation. Lower-scoring samples are refined by Level-1 annotators through iterative rewriting and re-evaluation (up to three iterations) before final validation. The annotation process involves three annotators (all co-authors), with clear role separation: two annotators handle refinement, and one performs final validation.





\subsection{Dataset Statistics}

\dataset contains 2,487 instances split into training, validation, and test sets following a 70/10/20 ratio. Each instance belongs to one of the trigger categories shown in Table~\ref{tab:dataset_distribution}.

\subsection{Task Formulation}

\dataset defines a conditional generation task:
\[
f(\texttt{trigger}, \texttt{context}) \rightarrow \texttt{opener}
\]
where the model maps a structured social context to a single-turn conversational initiation. Here, \textit{trigger} represents the intent to proactively initiate a conversation (e.g., medicine reminder, small talk, etc), while \textit{context} corresponds to the structured contextual schema.

\subsection{Use Case and Ethics}

\dataset is primarily intended for training and evaluating small language models capable of proactive, contextually calibrated conversation initiation. In this work, it is used to compare fine-tuned SLMs against zero-shot and large-scale models.

All data in \dataset is fully synthetic and contains no real user information or personally identifiable data. Health-related scenarios are included solely in synthetic form. The dataset is designed to support privacy-preserving, on-device conversational systems while ensuring generated content remains supportive, non-intrusive, and contextually appropriate.

\section{Model Fine-tuning}

\subsection{Training Paradigm}

We adopt a supervised fine-tuning (SFT) paradigm to train models for proactive conversational opener generation. Each training instance pairs a structured contextual representation with a target single-sentence opener. Models are optimized using a standard next-token prediction objective, minimizing cross-entropy loss over the reference output.

\subsection{Models}

To evaluate generalization across architectures, we fine-tune three instruction-aligned small language models -- \textbf{llama-3.2-3b-instruct}~\cite{grattafiori2024llama}, \textbf{gemma-3-4b-it}~\cite{team2024gemma}, and \textbf{phi-3.5-mini-instruct}~\cite{abdin2024phi3technicalreporthighly}. These models were selected to balance efficiency and capability while spanning diverse pretraining and alignment strategies.

\subsection{Input Representation}

All structured contextual information is serialized into a linearized plain-text format. Each input consists of an instruction followed by key–value pairs encoding the contextual schema and trigger information. This representation allows models to process heterogeneous contextual signals without architectural modification, while preserving interpretability and consistency across samples. The target output is a single, natural-language conversational opener. Model-specific chat templates are used to conform to each architecture’s instruction-following format.

\subsection{Parameter-Efficient Fine-Tuning}

We employ Low-Rank Adaptation (LoRA) \cite{hu2022lora} with 4-bit quantization (QLoRA) \cite{dettmers2023qlora} to enable efficient fine-tuning under limited computational resources. LoRA adapters are applied to key projection layers, allowing task-specific adaptation while keeping base model parameters frozen.

\begin{table*}[t]
\centering
\scriptsize
\setlength{\tabcolsep}{3pt}
\resizebox{\textwidth}{!}{
\begin{tabular}{l l c c c c c c c c c c}
\hline
\textbf{Model} & \textbf{Type} & \textbf{BLEU-1 $\uparrow$} & \textbf{BLEU-2 $\uparrow$} & \textbf{R-1 $\uparrow$} & \textbf{R-2 $\uparrow$} & \textbf{R-L $\uparrow$} & \textbf{BERT-F1 $\uparrow$} & \textbf{Embed $\uparrow$} & \textbf{PPL $\downarrow$} & \textbf{BLEURT $\uparrow$} & \textbf{COMET $\uparrow$} \\
\hline

llama-70b & Pre-trained & 0.0482 & 0.0248 & 0.1749 & 0.0454 & 0.1557 & 0.8612 & 0.4466 & 97.8867 & -1.0839 & 0.5193 \\
grok-4-1-fast  & Pre-trained & 0.1689 & 0.0955 & 0.2635 & 0.0892 & 0.2226 & 0.8768 & 0.5544 & 177.1497 & -0.9470 & 0.5843 \\
gemini-2.5-flash & Pre-trained & 0.1536 & 0.0895 & 0.2601 & 0.0938 & 0.2260 & 0.8804 & 0.5547 & 98.1497 & -0.7530 & 0.5878 \\
llama-3.2-3b-instruct & Pre-trained & 0.1239 & 0.0495 & 0.1892 & 0.0418 & 0.1529 & 0.8613 & 0.3783 & 59.6849 & -0.8925 & 0.5653 \\
phi-3.5-mini-instruct  & Pre-trained & 0.1609 & 0.0734 & 0.2215 & 0.0519 & 0.1799 & 0.8689 & 0.4195 & 68.5979 & -0.8835 & 0.5960 \\
gemma-3-4b-it  & Pre-trained & 0.0652 & 0.0270 & 0.1489 & 0.0309 & 0.1308 & 0.8561 & 0.3262 & 157.3717 & -1.1059 & 0.4985 \\

\hline

llama-3.2-3b-instruct  & Fine-tuned & 0.2734 & 0.1776 & 0.3463 & 0.1523 & 0.3020 & 0.8940 & 0.6151 & 59.7362 & -0.5086 & 0.6463 \\
phi-3.5-mini-instruct  & Fine-tuned & 0.2072 & 0.1178 & 0.3082 & 0.1064 & 0.2548 & 0.8849 & 0.5872 & 65.6999 & -0.5680 & 0.6354 \\
gemma-3-4b-it  & Fine-tuned & \textbf{0.2864} & \textbf{0.1888} & \textbf{0.3667} & \textbf{0.1685} & \textbf{0.3174} & \textbf{0.8962} & \textbf{0.6286} & \textbf{59.0560} & \textbf{-0.4606} & \textbf{0.6534} \\
\hline
\end{tabular}
}
\caption{Automatic evaluation results across pre-trained and fine-tuned models. Best scores are highlighted in bold.}
\label{tab:automatic_results}
\vspace{-2em}
\end{table*}

\subsection{Hardware Setup}

All fine-tuning experiments were conducted on a single consumer-grade GPU (NVIDIA RTX 4090, 16\,GB VRAM). The use of quantization-aware parameter-efficient fine-tuning enables training without specialized multi-GPU or server-class hardware, aligning with our goal of evaluating deployment-feasible SLMs.

\subsection{Training Considerations}

Models were fine-tuned using early stopping based on validation loss to prevent overfitting on the relatively compact dataset. All models were trained with identical optimization settings to ensure fair comparison across architectures. No additional data augmentation or prompt engineering was applied during fine-tuning.

\section{Evaluation}



Evaluating socially intelligent openers extends beyond conventional text generation tasks, as a given context may admit multiple valid and effective responses. Openers can vary in focus (e.g., emotion, environment, or task), strategy, and tone while remaining socially appropriate. Consequently, strict comparison against a single reference opener may underestimate the quality of diverse yet valid generations.

To address this, we adopt a three-pronged evaluation strategy combining similarity-based metrics with context-aware assessment methods.

\subsection{Automatic Metrics}
\textit{Evaluation Focus: How similar is the generated opener to the ideal opener?}

We employ a suite of automatic evaluation metrics to quantify the similarity between generated openers and their corresponding reference (gold) responses. These metrics capture complementary aspects of generation quality, including lexical overlap, semantic similarity, fluency, and diversity.

\subsubsection{Lexical Overlap Metrics}
We use BLEU ~\cite{papineni2002bleu} and ROUGE (ROUGE-1, ROUGE-2, ROUGE-L)~\cite{lin2004rouge} to measure n-gram overlap between generated and reference openers. These metrics provide a surface-level indication of how closely the generated text aligns with the reference in terms of wording and phrasing.

\subsubsection{Semantic Similarity Metrics}
To better capture meaning beyond exact word overlap, we employ BERTScore~\cite{zhang2019bertscore}, embedding-based cosine similarity using sentence transformers~\cite{reimers2019sentence}, BLEURT~\cite{sellam2020bleurt}, and COMET~\cite{rei2020comet}. These metrics leverage contextual embeddings and pre-trained models to assess whether the generated opener conveys similar intent and semantics as the reference, even when phrased differently.

\subsubsection{Fluency Metric}
We measure the fluency of generated openers using perplexity, computed with a pre-trained language model (gpt-2). Lower perplexity indicates more fluent and well-formed text.

Together, these metrics provide a comprehensive automatic evaluation of generation quality, balancing similarity to reference responses with diversity and fluency.

\subsection{Automatic Metrics Results}

Table~\ref{tab:automatic_results} presents the performance of all pre-trained and fine-tuned models across the automatic evaluation metrics.
Among all models, \textbf{gemma-3-4b-it (fine-tuned)} achieves the highest scores across all key metrics, including BLEU, ROUGE, BERTScore, embedding similarity, BLEURT, and COMET. This indicates that the model not only produces responses closer to the reference in terms of wording but also better captures the intended meaning and conversational quality. Additionally, its relatively low perplexity suggests improved fluency and coherence.

In contrast, pre-trained models exhibit significantly lower scores across all metrics. Among them, gemini-2.5-flash and grok-4-1-fast perform relatively better in semantic similarity metrics, such as BERTScore and embedding similarity, indicating a stronger inherent ability to capture meaning despite lower lexical overlap. phi-3.5-mini-instruct (pre-trained) and llama-3.2-3b-instruct (pre-trained) show moderate performance, while gemma-3-4b-it (pre-trained) and llama-70b (pre-trained) under perform on most metrics, particularly in overlap-based measures.

Perplexity results further highlight, fine-tuned models consistently achieve lower perplexity scores compared to their pre-trained counterparts, indicating more natural and coherent text generation.

In summary, automatic evaluation demonstrates that fine-tuning leads to substantial gains across all dimensions, with gemma-3-4b-it (fine-tuned) emerging as the strongest model overall. However, as discussed earlier, these metrics primarily capture similarity to reference responses and do not fully reflect the broader notion of social appropriateness, which we evaluate in subsequent sections.

\subsection{LLM-as-a-Judge Evaluation}

\textit{Evaluation Focus: How well does the generated opener align with the given context and social scenario?}

To complement automatic diversity metrics, we employ an \textit{LLM-as-a-Judge} evaluation framework to assess the quality of generated proactive conversational openers. This approach enables structured, context-aware evaluation of generation quality along multiple nuanced dimensions that are difficult to capture with surface-level metrics.

\paragraph{Evaluation Setup}
For each generated opener, the evaluator model is provided with:
\begin{itemize}[leftmargin=*]
\item the full structured context describing the user state, environment, memory, and trigger conditions,
\item the specific trigger under which the opener was generated,
\item the generated opener itself.
\end{itemize}

The evaluator is instructed to act as an expert judge and assess how effectively the opener utilizes the provided structured information while maintaining natural conversational quality. To improve robustness and reduce model-specific bias, we employ three independent state-of-the-art large language models as evaluators, namely GPT-5, GPT-5-mini, and Grok-4.3. All models independently score each generated opener under identical evaluation conditions, enabling a more reliable and balanced assessment across different evaluation perspectives.

\paragraph{Evaluation Objective}
The primary objective of this evaluation is to measure intelligent contextual prioritization. Specifically, the evaluator determines whether the generated opener can select relevant contextual variables, ignore irrelevant or noisy information, adapt appropriately to trigger-specific requirements, and maintain coherence and conversational naturalness

Importantly, the evaluation does not reward the use of all available variables. Instead, it emphasizes \textit{selective relevance}, where only contextually important information should influence the opener.

\paragraph{Evaluation Dimensions}
Each opener is evaluated across six dimensions adopting the mechanism suggested by Lin \textit{et al.}~\cite{lin2023llm}.

\begin{enumerate}[leftmargin=*]
    \item \textbf{Trigger Alignment}: Measures whether the opener behaves appropriately given the trigger and task state.
    
    \item \textbf{Saliency Selection}: Assesses whether the opener selects the most relevant variables while ignoring less relevant or irrelevant ones, reflecting effective contextual prioritization.
    
    \item \textbf{Constraint Adherence}: Evaluates logical consistency, ensuring the opener does not violate constraints, hallucinate facts, or contradict known context.
    
    \item \textbf{Value Sensitivity}: Measures how well the opener adapts tone and phrasing based on modifiers such as urgency, emotional state, or engagement level.
    
    \item \textbf{Noise Suppression}: Assesses the ability to avoid unnecessary or irrelevant contextual references that could degrade conversational quality.
    
    \item \textbf{Coherence and Naturalness}: Evaluates fluency, conciseness, and human-likeness of the opener.
\end{enumerate}

\paragraph{Scoring Protocol}
Each dimension is assigned a score between $0.0$ and $5.0$, with at most one decimal place. The evaluator outputs structured JSON containing scores for all six dimensions. To obtain an overall quality signal, we compute the average score across all dimensions for each opener. An example of an evaluated opener and its corresponding scores is shown below.

\noindent\textbf{Opener:} \textit{``Your follow-up with the doctor is coming up, 
take a moment to gather essentials and head out when ready.''}

\begin{itemize}
    \item Trigger Alignment: 4.5
    \item Saliency Selection: 4.0
    \item Constraint Adherence: 5.0
    \item Value Sensitivity: 4.0
    \item Noise Suppression: 5.0
    \item Coherence and Naturalness: 4.5
\end{itemize}

\noindent\textbf{Average Score:} 4.5

\paragraph{Quality Thresholding}
We define a high-quality threshold based on the average score of each opener. Openers with an average score $\geq 4.0$ are considered good openers, reflecting strong contextual reasoning, relevance, and conversational effectiveness. This threshold establishes a rigorous standard for success. By combining structured context understanding with multi-dimensional scoring, this method offers a comprehensive assessment of proactive conversational generation quality.

\begin{table}[t]
\centering
\resizebox{\columnwidth}{!}{%
\begin{tabular}{l | c c c | c}
\hline
\textbf{Model Name} & \textbf{GPT-5-mini} & \textbf{GPT-5} & \textbf{Grok-4.3} & \textbf{Avg.} \\
\hline
llama-70b$^1$ & 30.00 & 31.12 & 73.86 & 44.99 \\
grok-4-1-fast$^1$ & 69.00 & 64.94 & 77.39 & 70.44 \\
gemini-2.5-flash$^1$ & 60.58 & 59.54 & 80.50 & 66.87 \\
\hline
llama-3.2-3b-instruct$^2$ & 24.69 & 17.22 & 53.94 & 31.95 \\
phi-3.5-mini-instruct$^2$ & 46.27 & 40.04 & 75.10 & 53.80 \\
gemma-3-4b-it$^2$ & 27.59 & 25.10 & 54.36 & 35.68 \\
\hline
llama-3.2-3b-instruct$^3$ & 79.46 & 80.50 & \textbf{90.25} & 83.40 \\
phi-3.5-mini-instruct$^3$ & 76.14 & 70.54 & 85.68 & 77.45 \\
gemma-3-4b-it$^3$ & \textbf{80.29} & \textbf{81.33} & 89.21 & \textbf{83.61} \\
\hline
\end{tabular}}
\caption{LLM-as-judge evaluation of conversation openers: percentage of outputs classified as ``good'' by three judges (GPT-5 mini, GPT-5, Grok-4.3) and their average. model$^1$ - pre-trained LLMs, model$^2$ - pre-trained SLMs, and model$^3$ - fine-tuned SLMs.}
\label{tab:llm_judge_results}
\vspace{-2em}
\end{table}

\subsection{LLM-as-a-Judge Evaluation Results}

Table~\ref{tab:llm_judge_results} presents the results of the LLM-as-a-Judge evaluation, reporting the proportion of \textit{good openers} (average score $\geq 4.0$) as assessed by three independent evaluator models. We additionally report the average performance across evaluators to capture overall robustness.

\paragraph{Overall Performance Trends}
A consistent performance gap is observed between pre-trained models and fine-tuned models across all evaluators. Fine-tuned small language models (SLMs) achieve the strongest results overall, outperforming both their pre-trained counterparts and larger pre-trained language models. Importantly, this trend holds across all three judge models, indicating that the observed improvements are not tied to a specific evaluator but reflect general gains in generation quality.

Among the evaluated systems, fine-tuned variants of gemma-3-4b-it, llama-3.2-3b-instruct, and phi-3.5-mini-instruct consistently achieve the highest performance. In particular, \textbf{gemma-3-4b-it} demonstrates the strongest overall results when considering the average across evaluators, with closely competitive performance from the other fine-tuned models. This suggests that task-specific adaptation enables smaller models to match or exceed the capabilities of significantly larger models for this task.

\paragraph{Comparison with Pre-trained LLMs}
Among pre-trained large language models, grok-4-1-fast demonstrates the strongest performance, followed by gemini-2.5-flash and llama-70b. While these models generate a substantial number of high-quality openers, their performance remains below that of fine-tuned SLMs.

\paragraph{Pre-trained SLM Performance}
Pre-trained SLMs perform the weakest overall across all evaluators. While phi-3.5-mini-instruct shows comparatively stronger performance within this group, models such as llama-3.2-3b-instruct and gemma-3-4b-it struggle to produce high-quality openers without task-specific adaptation. This highlights the limitations of smaller models when applied in a zero-shot setting to tasks requiring nuanced contextual reasoning and socially appropriate language generation.

Overall, the results indicate that task-specific fine-tuning is a critical factor in enabling smaller models to achieve strong performance in proactive conversation initiation, while also providing robustness across diverse evaluation frameworks.

\subsection{Pairwise Evaluation}

\textit{Evaluation Focus: Which opener is better in a direct comparison under the same context?}

To further validate the effectiveness of fine-tuning, we conduct a \textit{pairwise evaluation} between the strongest pre-trained model and the best-performing fine-tuned model identified in the previous experiments.

\paragraph{Evaluation Setup}
We select grok-4-1-fast as the top-performing pre-trained model and gemma-3-4b-it (fine-tuned) as the best-performing fine-tuned model. For each shared context and trigger, both models generate an opener.

These paired openers are then evaluated by an LLM judge (GPT-5-mini), which is provided with:
\begin{itemize}
    \item the full structured context,
    \item the trigger condition,
    \item opener 1 (pre-trained),
    \item opener 2 (fine-tuned).
\end{itemize}

The evaluator assigns scores to each opener across the same six dimensions used in the standalone evaluation (Trigger Alignment, Saliency Selection, Constraint Adherence, Value Sensitivity, Noise Suppression, and Coherence). Additionally, the evaluator selects a \textit{winner} for each pair.
\begin{itemize}
    \item Opener 1 (pre-trained)
    \item Opener 2 (fine-tuned)
    \item Tie (if both are comparable)
\end{itemize}

\begin{table}[t]
\centering
\small
\begin{tabular}{l c c c}
\toprule
\textbf{Model} & \textbf{Avg. Score} & \textbf{Wins} & \textbf{Ties} \\
\midrule
grok-4-1-fast (pre-trained) & 4.1169 & 136 & \multirow{2}{*}{9} \\
gemma-3-4b-it  (fine-tuned) & \textbf{4.5539} & \textbf{337} & \\
\bottomrule
\end{tabular}
\caption{Pairwise evaluation between the best pre-trained and fine-tuned models. Wins indicate the number of times a model's opener was preferred by the LLM judge.}
\label{tab:pairwise_results}
\vspace{-2em}
\end{table}

\subsection{Pairwise Evaluation Results}

Table~\ref{tab:pairwise_results} summarizes the pairwise comparison results.
The results show a decisive advantage for the fine-tuned model. gemma-3-4b-it (fine-tuned) achieves a significantly higher average score (4.5539 vs. 4.1169) and is preferred in the majority of comparisons, winning 337 out of 482 cases. In contrast, grok-4-1-fast (pre-trained) wins only 136 cases, with 9 ties. This improvement is statistically significant (paired test, $p < 0.001$).

This corresponds to a win rate of approximately $70\%$ for the fine-tuned model, indicating a consistent and substantial improvement in opener quality.

The pairwise evaluation highlights several important findings.
\begin{itemize}
    \item Fine-tuned models not only improve absolute quality scores but also outperform strong pre-trained LLMs in direct comparison.
    \item The margin of improvement is consistent across diverse contexts, as reflected by the high number of wins.
    \item The low number of ties suggests that differences in quality are often clear and distinguishable.
\end{itemize}

These results reinforce the findings from the LLM-as-a-Judge evaluation, demonstrating that fine-tuning leads to more contextually appropriate, coherent, and effective proactive conversational openers. Even when compared against a strong pre-trained LLM, the fine-tuned model exhibits a clear and consistent advantage.


\section{Discussion and Conclusion}

\subsection{Key Findings}

Our evaluation demonstrates that fine-tuning small language models (SLMs) significantly improves the quality of proactive conversational openers. Across automatic metrics, fine-tuned models consistently outperform their pre-trained counterparts in fluency and semantic relevance, indicating improved internalization of contextual signals.

These results are corroborated by LLM-as-a-Judge evaluations, which capture higher-level conversational qualities beyond surface metrics. Fine-tuned SLMs substantially outperform both baseline SLMs and larger baseline LLMs, achieving success rates of up to 84\% compared to 70\% for the strongest baseline LLM and 54\% for baseline SLMs (Table~\ref{tab:llm_judge_results}). This suggests that targeted fine-tuning enhances contextual reasoning and conversational calibration even in compact models.

Pairwise preference evaluations provide strong evidence of qualitative improvement. The fine-tuned Gemma model outperforms the best-performing baseline model, winning approximately 70\% of head-to-head comparisons (Table~\ref{tab:pairwise_results}). This consistent preference indicates that context-driven fine-tuning yields conversational openers that are not only technically stronger but also perceptually superior.

\subsection{Limitations and Future Work}

This work focuses on single-turn conversation initiation, which does not fully capture real-world multi-turn interaction. Extending context-driven reasoning across longer dialogues, including adaptive memory updates and evolving user state, remains an important direction for future research. It is currently unclear to what extent models fine-tuned for initiation generalize to sustained multi-turn interaction without degrading conversational quality. Investigating whether a single model can both initiate and maintain dialogue, or whether specialised components are required, is an important avenue for future work.

Additionally, the approach relies on a curated set of structured context variables. While this enables controlled reasoning, generalization to more diverse or unstructured real-world contexts may require broader variable coverage and larger-scale data collection. A key challenge for real-world deployment lies in real-time extraction and grounding of these variables from sensor data. Future work will therefore explore integrating perception modules, such as vision and multimodal sensing systems, to infer contextual variables (e.g., user state, environment, and social cues) from raw sensory input. Addressing the reliability and latency of such pipelines will be essential for enabling fully autonomous proactive conversational agents in situated environments.

\subsection{Conclusion} 
Overall, this work shows that proactive conversation initiation benefits substantially from explicit context modeling and targeted fine-tuning. By combining a structured schema with a curated dataset, we demonstrate that compact, locally deployable language models can produce context-driven openers that match or exceed the performance of larger models. These findings highlight a promising direction for privacy-preserving, socially aware conversational agents.

\bibliographystyle{IEEEtran}
\bibliography{main}

\end{document}